\documentclass[twocolumn,letter]{jpsj3}
\usepackage{txfonts}
\usepackage{color}

\title{Weak-Light Nonlinearity Using a Dark State in Coupled Quantum Dots}

\author{Nobuhiko Yokoshi$^1$\thanks{yokoshi@pe.osakafu-u.ac.jp} and Hajime Ishihara$^{1,2}$}
\inst{$^1$Department of Physics and Electronics, Osaka Prefecture University, Sakai, Osaka 599-8531, Japan \\
$^2$Department of Materials Engineering Science, Osaka University,  Toyonaka, Osaka 560-8531, Japan} %\\

\abst{
We propose a scheme to induce weak-light nonlinearity in a double quantum dot. The scheme positively utilizes locality and dissipation of an external auxiliary system. As a plausible setup, we consider a complex system in which a localized plasmon field from a metallic nanotip couples with only one of the coupled quantum dots. The perturbative calculation with respect to the light intensity shows that, even by a sufficiently weak light, a dipole-forbidden two-exciton NOON state is prepared as the steady state. This result can be explained by combining the two factors: decoherence-induced quantum state preparation and two-photon resonance. The present work implies that the positive usage of both the locality and the dissipation in the external auxiliary system is promising for inducing two-photon processes effectively, and provides one guideline to weak-light nonlinearities. 
}
\begin{document}
\maketitle
Weak-light nonlinearity has been a major avenue of research in pure and applied physics. It is essentially required in establishing classical and quantum technologies, e.g., up-conversion for efficient solar cells~\cite{TTA-UC}, visible-to-telecom frequency conversion~\cite{Single-VTconv} and two-photon gateway for quantum communication~\cite{2photon-Gate}. However, nonlinear responses by weak light are considerably small. One of the plausible ways to overcome the problem is the usage of quantum interferences. There are various proposals using the inter-level interference inside four-level atoms~\cite{nonlinear0,nonlinear1,nonlinear2,nonlinear3,com-nonlinear1} and inter-cavity interference~\cite{Bamba2,Bamba}.  Actually, nonlinear optical effects due to such quantum interferences were experimentally achieved at a single photon level~\cite{nonlinear4,nonlinear5,nonlinear6,nonlinear7,nonlinear8}, and have contributed to a step for photon-based gateways and transistors.  

Here, we propose another scheme to realize a nonlinear excitation more simply with a sufficiently weak light. Its operating principle is based on the combination of the decoherence-induced quantum state preparation~\cite{DIEHL,Kraus,Stannigel} and the two-photon resonance. The essence of our scheme is the usage of an external auxiliary system, which is highly dissipated and interacts with only a part of target. The auxiliary system enhances the effective coupling between the light and target as well as attenuates all the excited states except an two-excitation NOON state~\cite{Sanders,Boto} that is moreover dipole-forbidden. It follows that only the surviving state determines the steady state. The aim of this work is to show that our scheme is functioning with proposing a plausible setup. 
\begin{figure}[b]
\begin{center}
\includegraphics[width=60mm]{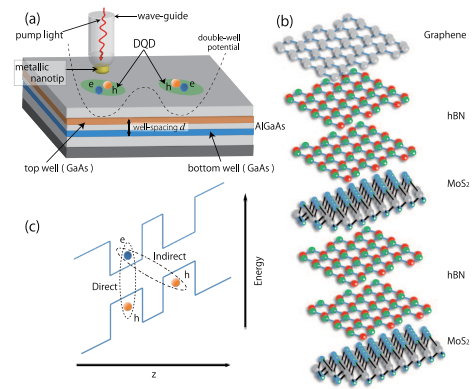}
\end{center}
\caption{(Color online) (a) An example of a DQD containing indirect excitons in coupled quantum wells. The DQD is coupled with a metallic nanotip attached on the top of a wave-guide. The applied base voltage confines the holes and electrons into the top and bottom quantum wells. The top gates create a double well potential for the excitons. (b) Similar system can be designed using van der Waals heterostructure, where, e.g., GaAs is replaced by MoS$_2$ monolayer and AlGaAs by an insulating hBN surrounded by cladding layers. (c) Schematic view of the direct and indirect excitons. Here $z$-axis is taken in the layer stacking direction.}
\label{model} 
\end{figure}

For that aim, we consider a gate-defined double quantum dot (DQD) in, e.g., semiconductor coupled quantum wells~\cite{Miller,Hammack,Remeika,Voros,Seidl,Kuznetsova,Tim,Schinner,Winbow,Asgari,Schinner2,Schinner3,Beian} or van der Waals heterostructure~\cite{vW1,vW2,vW3} [see Figs.~\ref{model}(a) and~\ref{model}(b)]. In such systems, one can electrostatically confine indirect excitons by quantum confined Stark effect. Because of the spatial separation between the electron and hole layers [Fig.~\ref{model}(c)], the lifetime of the indirect excitons reaches typically from 10ns to 10$\mu$s~\cite{Voros,Seidl,Asgari}, and the excitation energies of them can be controlled by the back gate potential~\cite{Hammack,Schinner,Winbow,Schinner2}. In Ref. [32], the separation between the quantum wells is $d=17$nm and the Bohr radius is estimated to be $a_B=18$nm, whereas the confinement energy is $\hbar \Omega_c \sim 0.8$meV. Considering such a situation, we can safely approximate the indirect excitons as soft-core bosons~\cite{Tim}. Actually, single exciton emissions were observed from two- and three-exciton states in the parabolically-confined quantum dot~\cite{Schinner3}.

\begin{figure}[bt]
\begin{center}
\includegraphics[width=50mm]{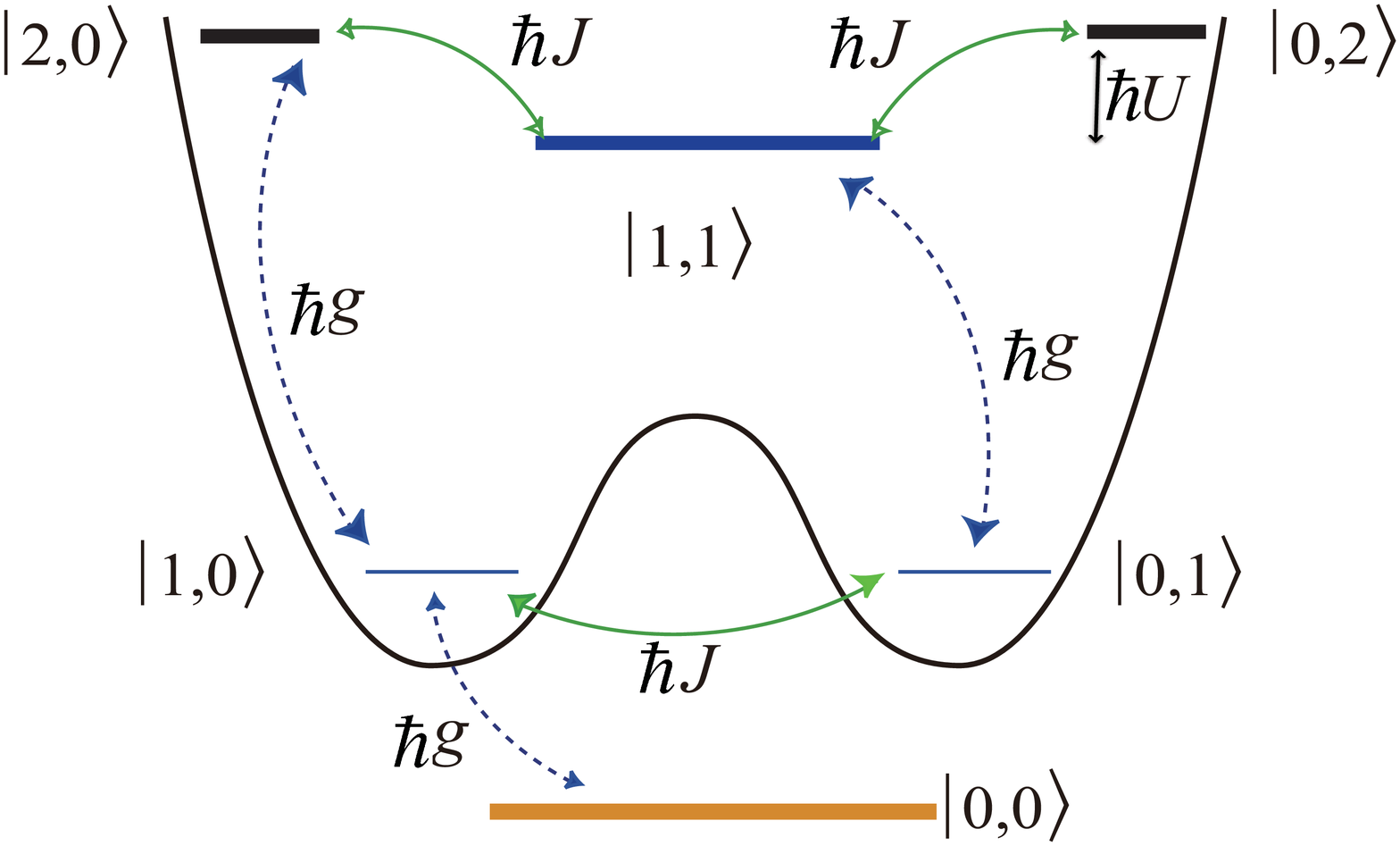}
\end{center}
\caption{(Color online) Schematic view of the scattering processes and the energy diagram in the DQD. The inter-dot hopping energy $\hbar J$ lifts the degeneracy of the singly and doubly occupied states of the indirect excitons. The nonlinearity in the optical excitations of the excitons is described through the dipole-dipole interactions $\hbar U$ and $\hbar X$. Because of the locality of the plasmon-enhanced electric field, the auxiliary plasmon couples with only one of the quantum dots. The plasmon-dot coupling energy is $\hbar g$.}
\label{diagram} 
\end{figure}

In the DQD, the number of the light-pumped excitons in each dot characterizes the Hilbert space. We label the states by $|m,n\rangle$ with $m,n$ being integers. The extended Hubbard model for the indirect excitons is written as 
\begin{eqnarray}
&&H_d=\sum_{i=1,2} \bigl( \hbar\varepsilon b_i^{\dagger}b_i + \frac{\hbar U}{2}b_i^{\dagger}b_i^{\dagger}b_i b_i \bigr)
\nonumber \\
&&~~~~~~~~~~~
+\bigl( \hbar J b_1^{\dagger} b_2 + \hbar X b_1^{\dagger} b_2^{\dagger} b_2 b_1 + {\rm H.c.}\bigr),
\end{eqnarray}
where $b_i$ $(b_i^{\dagger})$ is the annihilation (creation) operator in $i$-th quantum dot. Throughout this work, the spin of the excitons is disregarded because we assume that a polarized light excites only one of the spin species~\cite{Tim}. The excitation energy $\hbar \varepsilon$ and inter-dot hopping $\hbar J<0$ can be controlled by gate potentials (see Fig.~\ref{diagram} )~\cite{Hammack,Schinner,Winbow,Schinner2}. The nonlinearity of the exciton excitation is introduced through the on-site and inter-site dipole-dipole interactions. The on-site dipole-dipole interaction $\hbar U$ depends on the separation $d$, and is always repulsive~\cite{Tim}. The interaction is estimated from the parameters to be $\hbar U \sim 2$meV~\cite{Schinner3}. The inter-dot interaction $\hbar X$ is typically smaller than $\hbar U$. The radiative decay of the indirect excitons is described in the conventional Lindblad form as
\begin{eqnarray}
{\mathcal L}\rho(t)=\frac{\hbar \gamma}{2}\sum_{i=1,2} \Bigl[ 2b_i \rho(t) b_i^{\dagger}-b_i^{\dagger}b_i\rho(t)-\rho(t)b_i^{\dagger}b_i \Bigr],
\end{eqnarray}
where the radiative decay energy  $\hbar \gamma \sim 0.04 \mu$eV~\cite{Schinner3}. The density matrix of the system $\rho(t)$ is defined below.

As for the auxiliary system, we use an metallic nanotip located at the top of a wave-guide~\cite{Anger,Berweger}. Although the strong dissipation in the metal can be a serious bottleneck for weak-light nonlinearity~\cite{antenna1,antenna2}, it is known that the quantum interference between the plasmon and target provides a positive insight into the solutions~\cite{Osaka1,Osaka2,Ishihara,Tanaka,ET}. When the typical size of the metallic structures is much smaller than the light wavelength, they exhibit the saturations in the optical absorption and emission, which can be qualitatively explained within the two-level plasmon model~\cite{Chu,Mohan,Elim,HataJ,Hata}. Thus, we employ the Hamiltonian for the plasmons as 
\begin{eqnarray}
H_p= \hbar \omega_p (\sigma_+\sigma_- -\frac{1}{2}) +\hbar\kappa (\sigma_+ e^{-i\omega_L t}+\sigma_- e^{i\omega_L t}),
\end{eqnarray}
where $\sigma_{\pm}$ is the ladder operator of the two-level plasmon separated by the frequency $\omega_p$, and $\kappa$ is the pump rate of the monochromatic light with the frequency $\omega_L$. Because the spatial spread of the plasmon-induced electric field is restricted to the several tens of nanometer, we consider the situation that the auxiliary plasmon field couples with only one of the quantum dots [see Fig.~\ref{model}(a) and Fig.~\ref{diagram}]. The corresponding interaction Hamiltonian is 
 \begin{eqnarray}
 H_{I}=\hbar g(b_1 \sigma_+ +b_1^{\dagger} \sigma_-).
 \end{eqnarray}
It follows that the pump light simultaneously excites both the superposition states $|\Phi^{1-}\rangle \equiv ( |1,0;0\rangle - |0,1;0\rangle)/\sqrt{2}$ and $|\Phi^{1+}\rangle \equiv ( |1,0;0\rangle + |0,1;0\rangle)/\sqrt{2}$ by exciting the state $|1,0;0\rangle$ through the localized plasmon. As for the case in the indirect excitons, we treat the radiative decay and dephasing in the Lindblad form~\cite{HataJ}
\begin{eqnarray}
{\mathcal L}\rho(t)=\frac{\hbar \Gamma_r}{2}\sum_{i=1,2} \Bigl[ 2\sigma_- \rho(t) \sigma_+ -\sigma_+\sigma_-\rho(t)- \rho(t)\sigma_+\sigma_- \Bigr], \\
{\mathcal L}\rho(t)=\hbar \Gamma_{\phi}\sum_{i=1,2} \Bigl[ \sigma_z \rho(t) \sigma_z -\rho(t) \Bigr],
\end{eqnarray}
where $\sigma_z=(\sigma_+\sigma_- -1/2)$ and the parameter $\Gamma_r$ and $\Gamma_{\phi}$ are the radiative decay rate and dephasing rate of the two-level plasmon. The total dissipation $\Gamma=\Gamma_r+\Gamma_{\phi}$ is typically much larger than the other coupling constants. 

We briefly mention the intensity of the pump light. From now on, we move to rotating frame of reference with respect to the pump light, i.e., $(\varepsilon-\omega_L) \rightarrow \varepsilon$ and $(\omega_p-\omega_L) \rightarrow \omega_p$. According to Fermi's golden rule, $\mathcal{N} \sim \kappa^2/g\Gamma$ photons interact with the metal tip during one Rabi cycle on average. We have set the parameters as $\hbar \kappa =\hbar g=0.2$meV and $\hbar \Gamma=20$meV, which result in $\mathcal{N}=0.01$. In addition, seeing the isolated metal tip ($g=0$), the excited state population of the plasmon becomes in the steady state
 \begin{eqnarray}
\langle \sigma_+\sigma_- \rangle=\frac{\kappa^2}{2\kappa^2 +(\Gamma/2)^2+\omega_p^2} \lessapprox 4.0\times 10^{-4}.
 \end{eqnarray}
Therefore, we consider that the whole system is sufficiently in few photon regime. 

Assuming the above system setup, we show that the positive usage of both the locality and the dissipation in the auxiliary plasmon is promising for inducing two-exciton excitation effectively. Before proceeding to the detailed calculations, we will outline the mechanism for such a nonlinear response. The large dissipation in the metallic tip essentially disturbs the coherence of the whole system, and attenuates also the exciton states with time. However, the localized access to the DQD enables us to make a two-exciton dark eigenstate that does not couple with the plasmon. Therefore, only the dark eigenstate survives in the steady state, because it is free from the disturbance from the plasmon. The steady state is so-called NOON state~\cite{Sanders,Boto}, and moreover its parity is odd with respect to the dot position. Thus, it is dipole-forbidden and does not relax to the ground state directly. This aspect prolongs the already long lifetime of the indirect excitons ($\lessapprox 10 \mu$s)~\cite{Voros,Seidl,Asgari}. Therefore, we can obtain the quantum two-exciton state that is quite suited to manipulate coherently.

\begin{figure}[b]
\begin{center}
\includegraphics[width=70mm]{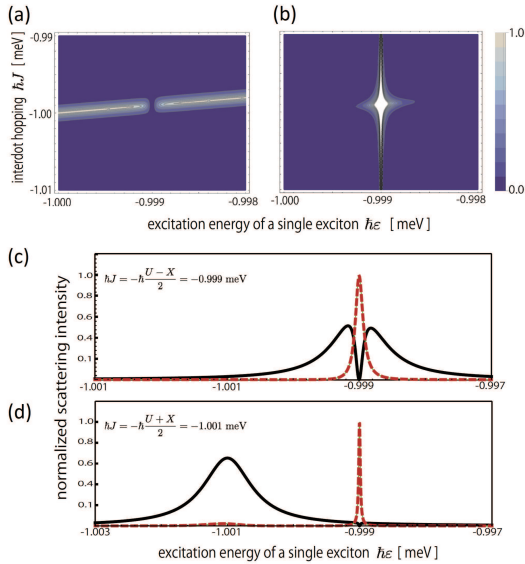}
\end{center}
\caption{(Color online) (a) Normalized scattering intensity of the state $|\Phi^{1-}\rangle$ in the steady state, that is given by $P^{1-}=|\alpha_1(\infty)-\alpha_2(\infty)|^2/2$, is plotted against the inter-dot coupling $\hbar J$ and the excitation energy $\hbar \varepsilon$. (b) The same plot of the intensity for the state $|\Phi^{2-}\rangle$, that is given by $P^{2-}=|\alpha_4(\infty)-\alpha_5(\infty)|^2/2$. (c) The sectional view of the scattering intensities at $J=-(U-X)/2$. The solid lines represent the intensity of $|\Phi^{1-}\rangle$, whereas the dashed ones represent the intensity of $|\Phi^{2-}\rangle$. (d) The same plot at $J=-(U+X)/2$. Here we have set the parameters as $\hbar \Gamma =20$meV, $\hbar \gamma =0.04 \mu$eV, $\hbar \kappa =0.2$meV, $\hbar g=0.2$meV, $\hbar U =2.0$meV, and $\hbar X =2.0 \mu$eV.}
\label{population} 
\end{figure}

To calculate the steady state, we employ the perturbation method developed by Carmichael {\it et al.}~\cite{Carmichael}. Let us consider a pure state $|\psi (t)\rangle$ which consists $|m,n; s\rangle \equiv |m,n\rangle \otimes | s\rangle$. Here $ | s\rangle$ is the two-level plasmon state with $s=\{ 0 ,1 \}$ denoting the ground and excited states. One derives the Liouville master equation with respect to $\rho(t) \equiv |\psi\rangle\langle\psi|$, and solves it perturbatively with respect to the ratio $\kappa/\Gamma$. Therefore, we take into account up to the two-excitation states ($\delta \equiv m+n+s \leq 2$). Therefore, we write down the pure state as follows
\begin{eqnarray}
&&|\psi(t)\rangle =
 \alpha_0(t)|0,0;0\rangle+\alpha_1(t)|1,0;0\rangle+\alpha_2(t)|0,1;0\rangle
\nonumber \\
&&~~~~~~~~
+\alpha_3(t)|1,1;0\rangle+\alpha_4(t)|2,0;0\rangle+\alpha_5(t)|0,2;0\rangle
\nonumber \\
&&~~~~~~~~
+\beta_0(t)|0,0;1\rangle+\beta_1(t)|1,0;1\rangle+\beta_2(t)|0,1;1\rangle.
\end{eqnarray}
Owing to the large dissipation of the plasmon ($\Gamma \gg \kappa$), we assume that the coefficient $\alpha_0(t)$ is independent of time~\cite{Carmichael}. In Figs.~\ref{population}(a-d), we show the normalized scattering intensities $\{ |\alpha_i|^2, |\beta_i|^2 \}$ in the steady state which are plotted against the excitation energy $\hbar\varepsilon$ and the inter-dot coupling $\hbar J$. It is found that the state $|\Phi^{1-}\rangle$ is well-excited centering around $\hbar \varepsilon = \hbar J$, whereas the state $|\Phi^{1+}\rangle$ does not appear. In addition, the scattering intensity for the two-exciton NOON state 
 \begin{eqnarray}
 |\Phi^{2-}\rangle \equiv  \frac{1}{\sqrt{2}} (|2,0;0\rangle-|0,2;0\rangle),
\end{eqnarray}
has a sharp peak at $\hbar \varepsilon = -\hbar (U-X)/2$. The intensities of the other excited states are not plotted in the figure because they are negligibly small. When the condition $\hbar J=-\hbar (U-X)/2$ is satisfied, the appearance of the two-exciton NOON state becomes robust against the deviation of the exciton energy from $\hbar \varepsilon =-\hbar (U-X)/2$. On the other hand, such a situation makes the scattering intensity of the one-exciton state suppressed in that regime [see Figs.~\ref{population}(c) and~\ref{population}(d)].

One may consider that the steady state should be more cluttered in the presence of the radiative decay and dephasing. However, the results in Fig.~\ref{population} can be rather understood qualitatively as being based on the scheme that positively utilizes such dissipations, i.e., decoherence-induced quantum state preparation~\cite{DIEHL,Kraus,Stannigel}. In order to discuss this consideration we utilize the following non-hermitian Hamiltonian:
 \begin{eqnarray}
 H_{\rm nh}= H_d+H_p+H_I-i\frac{\hbar \Gamma}{2}\sigma_+\sigma_- -i\frac{\hbar \gamma}{2} \sum_{i=1,2} b_{i}^{\dagger}b_i.
 \end{eqnarray} 
When this Hamiltonian has zero as an eigenenergy, there is always a dark state for the plasmon. The eigenstates, which include the plasmon-excited states $|m,n;1\rangle$, should have complex eigenenergies. Due to the imaginary part of the eigenenergy, such states are attenuated in time. In the beginning, let us consider the eigenenergies within the subspace \{$ |0,0;0\rangle$,$ |0,0;1\rangle$,$ |1,0;0\rangle$,$ |0,1;0\rangle$\}. When the condition $J = \varepsilon$ is satisfied, one of the eigenenergies actually becomes zero, to which the eigenstate $|D^1\rangle =\sin\theta |\Phi^{1-} \rangle+\cos\theta |0,0;0\rangle+ \mathcal{O}(\gamma)$ corresponds. Here we have defined the angle $\theta=-\arctan(\sqrt{2}\kappa/g)$. The other eigenstates include the plasmon-excited state $|0,0;1\rangle$, and their eigenenergies have the large imaginary parts due to the strong dissipation $\Gamma$. Because these states should decay promptly, the steady state is determined only by the dark state $|\Phi^{1-} \rangle$, which agrees with the result by the perturbative calculation. 

Drawing on the above consideration, we will move on to the consideration of the two-exciton states. In order to investigate the two-exciton dark state, we start by searching for the two-photon resonance. Within the two-exciton subspace $\{ |1,1;0\rangle, |2,0;0\rangle, |0,2;0\rangle \}$, the determinant of the Hamiltonian $H_d$ becomes
\begin{eqnarray}
\mathcal{D}=2\varepsilon \Upsilon^2  - 4J^2 (2\varepsilon+U-X),
\end{eqnarray}
where $\Upsilon^2 = (2\varepsilon+U)^2-X^2$. The condition for the two-photon resonance can be found by solving the equation $\mathcal{D}=0$ with respect to the excitation energy $\varepsilon$. Two of the solutions $\varepsilon=-(U+X\pm\sqrt{(U+X)^2+16J^2})/4$ are the resonance conditions to the states that consist of $|1,1;0\rangle$ and $|\Phi^{2+}\rangle\equiv (|2,0;0\rangle+|0,2;0\rangle)/\sqrt{2}$. These states are then dipole-allowed. The last one $\varepsilon= -(U-X)/2$ provides us with the resonance condition to the dipole-forbidden state $|\Phi^{2-}\rangle$. Note that the last solution does not depend on the inter-dot hopping $J$. Thus we can make it compatible with the condition for the efficient one-exciton excitation, i.e., $J = \varepsilon$.

Once the excitation energy $\hbar \varepsilon$ is controlled so that $\sqrt[3]{\mathcal{|D|}} \ll \kappa$, the two-exciton states become near resonant to the two-photon absorption. Thus their contributions cannot be neglected. Therefore, the state $|D^1\rangle$ becomes far from the eigenstate of the non-hermitian Hamiltonian $H_{\rm nh}$. Unfortunately, no dark state exists for the plasmon when we incorporate all the states with $\delta=0,1,2$. However, when the on-site interaction is much larger than the plasmon-dot coupling $(U \gg g)$ and the two-photon absorption is near resonant $(\varepsilon \sim -(U-X)/2)$, we can disregard the state $|\Phi^{1+}_e \rangle \equiv (|1,0;1\rangle+|0,1;1\rangle)/\sqrt{2}$. On this occasion, the state
\begin{eqnarray}
|D^2\rangle = \cos \varphi |D^1\rangle + \sin \varphi |\Phi^{2-}\rangle+ \mathcal{O}(\gamma),
\end{eqnarray}
can be regarded as an approximate dark state. Here the rotating angle is $\varphi=-\arctan(2\kappa\sin\theta/g)$. The other eigenstates essentially include the plasmon-excited states $|m,n;1\rangle$, and are attenuated with time. Therefore, through the surviving state $|D^2\rangle$, we can readily prepare the two-exciton NOON state $|\Phi^{2-}\rangle$ as the steady state. 
 
Subsequently, we mention the feasibility of the above nonlinear excitation. Fortunately, our scheme does not require strict conditions for the metallic nanotip. In the coupled quantum wells, the single exciton energy $\hbar \varepsilon$ is controlled for the range over 40 meV~\cite{Schinner2,Schinner3}. It is inter-dot hopping $J$ that we have to adjust more precisely. The incident light efficiently excites $|\Phi^{2-}\rangle$ for $J= -(U-X)/2$ (see Fig.~\ref{population}(c)). When the inter-dot hopping is increased, e.g., to be $J=-(U-X)$, the scattering intensities of the doubly occupied states drop down to $|\alpha_{4}|^2 \sim |\alpha_{5}|^2 \sim 2\times 10^{-3}$ even at the peak. Using the state-of-the-art technologies, the localization of the excitons is continuously controlled~\cite{Remeika}. Thus, one can construct the system setup with required precision. Another factor that impedes our scheme is the fluctuation of the exciton spins. Because we have considered the spin-polarized excitons, the spin fluctuations disturb the coherence of the steady NOON state. Fortunately, the spin fluctuation of the indirect excitons is small with a time comparable to the radiative lifetime~\cite{Seidl}. The spin lifetime is further increased when the excitons are spatially localized~\cite{Beian}. Therefore, in the quantum dot with a sub-micron wide electrostatic trap~\cite{Schinner3}, we believe that the proposed scheme has sufficient feasibility.

In conclusion, we have proposed a scheme to prepare a two-indirect exciton state in a DQD by sufficiently weak light. In the scheme, we simultaneously access both dipole-allowed and forbidden states in the DQD by utilizing localized electric field from a metallic nanotip. Although the usage of metallic structures with large dissipation seemingly has a worse hand for quantum state preparations, we have shown that only the dipole-forbidden NOON state can be left in the steady state by just waiting for a certain time. We found that this result is understood by considering dark states with respect to the dissipation in the nanotip. The resultant NOON state is expected to have much longer lifetime than that of the original indirect excitons  ($\sim 10 \mu$s) due to its parity, and thus quite suited to the field for quantum state manipulations. At last, the scheme may be applied also in multi-quantum dot systems. For instance, in the case where a single indirect exciton is excited in $N$-quantum dot ring, a dark state appears when a certain integer $k$ exists satisfying the relation $k = \frac{N}{2\pi}(\arccos(\varepsilon/2|J|)+\pi)$. The corresponding eigenstate is 
$|k\rangle=\frac{1}{\sqrt{N}} \Sigma_{m=1}^N\exp[i \frac{2\pi}{N} (k-\frac{N}{2}) m] b_m^{\dagger} |0,\cdots,0; 0\rangle.$
  If the result is extended to multi-exciton system in future, we can expect additional progress in applications to optics and quantum information technologies. 

This work was supported by JSPS KAKENHI Grant No. JP16H06504 in Scientific Research on Innovative Areas ``Nano-Material Optical-Manipulation'', Grant-in-Aid for Challenging Exploratory Research No. 15K13505.


\begin{thebibliography}{0} 
\bibitem{TTA-UC}%
S. Baluschev, T. Miteva, V. Yakutkin, G. Nelles, A. Yasuda, and G. Wegner, 
Phys. Rev. Lett. {\bf 97}, 143903 (2006).
\bibitem{Single-VTconv}%
S. Zaske, A. Lenhard, C. A. Ke{\ss}ler, J. Kettler, C. Hepp, C. Arend, R. Albrecht, W.-M. Schulz, M. Jetter, P. Michler, and C. Becher, 
Phys. Rev. Lett. {\bf 109}, 147404 (2012).
\bibitem{2photon-Gate}%
A. Kubanek, A. Ourjoumtsev, I. Schuster, M. Koch, P. W. H. Pinkse, K. Murr, and G. Rempe, 
Phys. Rev. Lett. {\bf 101}, 203602 (2008).
\bibitem{nonlinear0}%
H. Schmidt and A. Imamogdlu,
Opt. Lett. {\bf 21}, 1936 (1996).
\bibitem{nonlinear1}%
A. Imamoglu, H. Schmidt, G. Woods, and M. Deutsch,
Phys. Rev. Lett. {\bf 79}, 1467 (1997).
\bibitem{nonlinear2}%
S. E. Harris and Y. Yamamoto,
Phys. Rev. Lett. {\bf 81}, 3611 (1998).
\bibitem{nonlinear3}%
M.D. Lukin and A. Imamoglu,
Phys. Rev. Lett. {\bf 84}, 1419 (2000).
\bibitem{com-nonlinear1}%
M Bajcsy, A Majumdar, A Rundquist and J Vu$\check{\rm c}$kovi\'c,
New J. Phys. {\bf 15}, 025014 (2013).
\bibitem{Bamba2} T. C. H. Liew and V. Savona,
Phys. Rev. Lett. {\bf 104}, 183601 (2010).
\bibitem{Bamba} M. Bamba, A. Imamoglu, I. Carusotto, and C. Ciuti, 
Phys. Rev. A, {\bf 83}, 021802(R) (2011).
\bibitem{nonlinear4}%
K. J. Resch, J. S. Lundeen, and A. M. Steinberg,
Phys. Rev. Lett. {\bf 87}, 123603 (2001).
\bibitem{nonlinear5}%
K. M. Birnbaum, A. Boca, R. Miller, A. D. Boozer, T. E. Northup, and H. J. Kimble,
Nature {\bf 436}, 87 (2005).
\bibitem{nonlinear6}%
I. Schuster, A. Kubanek, A. Fuhrmanek, T. Puppe, P. W. H. Pinkse, K. Murr, and G. Rempe,
Nat. Phys. {\bf 4}, 382 (2008).
\bibitem{nonlinear7}%
J. Hwang, M. Pototschnig, R. Lettow, G. Zumofen, A. Renn, S. G\"otzinger, and V. Sandoghdar,
Nature {\bf 460}, 76 (2009).
\bibitem{nonlinear8}%
T. Peyronel, O. Firstenberg, Q.-Y. Liang, S. Hofferberth, A. V. Gorshkov, T. Pohl, D. Lukin, and Vladan Vuleti$\check{\rm c}$,
Nature {\bf 488}, 57 (2012).
\bibitem{DIEHL}
S. Diehl, A. Micheli, A. Kantian, B. Kraus, H. P. B\"ucher, and P. Zoller,
Nat. Phys. {\bf 4}, 878  (2008).
\bibitem{Kraus}
B. Kraus, H. P. B\"chler, S. Diehl, A. Kantian, A. Micheli, and P. Zoller,
Phys. Rev. A {\bf 78}, 042307 (2008).
\bibitem{Stannigel}
K. Stannigel, P. Rabl, and P. Zoller,
New J. Phys. {\bf 14}, 063014 (2012).
\bibitem{Sanders} 
B. C. Sanders, 
Phys. Rev. A {\bf 40}, 2417 (1989).
\bibitem{Boto} 
A. N. Boto, P. Kok, D. S. Abrams, S. L. Braunstein, C. P. Williams, and J. P. Dowling, 
Phys. Rev. Lett. {\bf 85}, 2733 (2000).
\bibitem{Miller} 
D. A. B. Miller, D. S. Chemla, T. C. Damen, A. C. Gossard, W. Wiegmann, T. H. Wood, and C. A. Burrus, 
Phys. Rev. B {\bf 32}, 1043 (1985).
\bibitem{Voros} 
Z. V\"or\"os, D. W. Snoke, L. Pfeiffer and K. West
Phys. Rev. Lett. {\bf 103}, 016403 (2009). 
\bibitem{Seidl} 
K. Kowalik-Seidl, X. P. V\"ogele, B. N. Rimpfl, S. Manus, J. P. Kotthaus, D. Schuh, W. Wegscheider, and A. W. Holleitner,
Appl. Phys. Lett. {\bf 97}, 011104 (2010).
\bibitem{Asgari} 
A. Asgari, S. Safa, and L. Mouchliadis,
Superlattices and Microstructures {\bf 49}, 487 (2011).
\bibitem{Hammack} 
A. T. Hammack, N. A. Gippius, G. O. Andreev, L. V. Butov, M. Hanson, and A. C. Gossard,
J. Appl. Phys. {\bf 99}, 066104 (2006).
\bibitem{Schinner} 
G.J. Schinner, E. Schubert, M.P. Stallhofer, J.P. Kotthaus, D. Schuh, A.K. Rai, D. Reuter, A.D. Wieck, A.O. Govorov, and J.P. Kotthaus, 
Phys. Rev. B {\bf 83}, 165308 (2011).
\bibitem{Winbow} 
A. G. Winbow, J. R. Leonard, M. Remeika, Y. Y. Kuznetsova, A. A. High, A. T. Hammack, L. V. Butov, J. Wilkes, A. A. Guenther, A. L. Ivanov, M. Hanson, and A. C. Gossard,
Phys. Rev. Lett. {\bf 106}, 196806 (2011).
\bibitem{Schinner2} 
G. J. Schinner, J. Repp, E. Schubert, A. K. Rai, D. Reuter, A. D. Wieck, A. O. Govorov, A. W. Holleitner, and J. P. Kotthaus,
Phys. Rev. B {\bf 87}, 205302 (2013).
\bibitem{Remeika}
M. Remeika, J. C. Graves, A. T. Hammack, A. D. Meyertholen, M. M. Fogler, and L. V. Butov, M. Hanson and A. C. Gossard,
Phys. Rev. Lett. {\bf 102}, 186803 (2009). 
\bibitem{Kuznetsova} 
Y. Y. Kuznetsova, A. A. High, and L. V. Butov,
Appl. Phys. Lett. {\bf 97}, 201106 (2010).
\bibitem{Tim} 
T. Byrnes, P. Recher, and Y. Yamamoto,
Phys. Rev. B {\bf 81}, 205312 (2010).
\bibitem{Schinner3} 
G. J. Schinner, J. Repp, E. Schubert, A. K. Rai, D. Reuter, A. D. Wieck, A. O. Govorov, A. W. Holleitner, and J. P. Kotthaus,
Phys. Rev. Lett. {\bf 110}, 127403 (2013).
\bibitem{Beian} 
M. Beian, M. Alloing, E. Cambril C. G. Carbonell, J. Osmond, A. Lema\^itre, and F. Dubin, 
Europhys. Lett. {\bf 110}, 27001 (2015).
\bibitem{vW1} 
F. Ceballos, M. Z. Bellus, H.-Y. Chiu, and H. Zhao,
ACS Nano {\bf 8}, 12717 (2014).
\bibitem{vW2} 
E. V. Calman, C. J. Dorow, M. M. Fogler, L. V. Butov, S. Hu, A. Mishchenko, and A. K. Geim,
Appl. Phys. Lett. {\bf 108}, 101901 (2016).
\bibitem{vW3} 
B. Miller, A. Steinhoff, B. Pano, F. Jahnke, A. Holleitner, U. Wurstbauer,
arXiv:1703.09566v1.
\bibitem{Anger} 
P. Anger, P. Bharadwaj, and L. Novotny,
Phys. Rev. Lett. {\bf 96}, 113002 (2006).
\bibitem{Berweger} 
S. Berweger, J. M. Atkin, R. L. Olmon, M. B. Raschke, J. Phys. Chem. Lett. {\bf 1}, 3427 (2010).
\bibitem{antenna1}%
P. Bharadwaj, B. Deutsch, and L. Novotny, 
Adv. Opt. Photon.  \textbf{1}, 438 (2009).
\bibitem{antenna2}%
L. Novotny and N. van Hulst, 
Nat. Photo. {\bf 5}, 83 (2011).
\bibitem{Ishihara} M. Nakatani, A. Nobuhiro, N. Yokoshi, and H. Ishihara, 
Phys. Chem. Chem. Phys. {\bf 15}, 8144 (2013). 
\bibitem{Osaka1} Y. Osaka, N. Yokoshi, M. Nakatani, and H. Ishihara, 
Phys. Rev. Lett. {\bf 112}, 133601 (2014).
\bibitem{Osaka2} Y. Osaka, N. Yokoshi, and H. Ishihara, 
Phys. Rev. B {\bf 93}, 155420 (2016)
\bibitem{ET}%
T. Utikal, T. Zentgraf, T. Paul, C. Rockstuhl, F. Lederer, M. Lippitz, and H. Giessen, 
Phys. Rev. Lett.  \textbf{106}, 133901 (2011).
\bibitem{Tanaka} 
Y. Tanaka, H. Ishiguro, H. Fujiwara, Y. Yokota, K. Ueno, H. Misawa, and K. Sasaki, 
Opt. Express {\bf 19}, 7726 (2011).
\bibitem{Elim} 
H. I. Elim, J. Yang, J.-Y. Lee, J. Mi, and W. Ji,
App. Phys. Lett. {\bf 88}, 083107 (2006).
\bibitem{Mohan} 
S. Mohan, J. Lange, H. Graener, and G. Seifert,
Opt. Exp. {\bf 20}, 28655 (2012).
\bibitem{Chu} 
S.-W. Chu, T.-Y. Su, R. Oketani, Y.-T. Huang, H.-Y. Wu, Y. Yonemaru, M. Yamanaka, H. Lee, G.-Y. Zhuo, M.-Y. Lee, S. Kawata, and K. Fujita,
Phys. Rev. Lett. {\bf 112}, 017402 (2014).
\bibitem{HataJ}
R. Hata, N. Yokoshi, H. Ajiki, and H. Ishihara,
J. Phys. Soc. Jpn. {\bf 83}, 093401 (2014).
\bibitem{Hata}
R. Hata, H. Ajiki, N. Yokoshi, and H. Ishihara,
Phys Status Solidi C {\bf 13}, 105 (2016).
\bibitem{Carmichael} H. J. Carmichael, R. J. Brecha, and P. R. Rice, 
Opt. Commun. {\bf 82}, 73 (1991).
\end{thebibliography}
\end{document}